\documentclass[%
 aip,
 amsmath,amssymb,
 reprint,%
]{revtex4-1}

\usepackage{graphicx}
\usepackage{dcolumn}
\usepackage{bm}

\usepackage[utf8]{inputenc}
\usepackage[T1]{fontenc}
\usepackage{mathptmx}
\usepackage{float}
\usepackage{color}

\newcounter{lastnote}

\begin{document}

\preprint{AIP/123-QED}

  \title{Active control of transport through nanopores}

\author{Cheng Lian}
 \affiliation{State Key Laboratory of Chemical Engineering, Shanghai Engineering Research Center of Hierachical Nanomaterials, and School of Chemistry and Molecular Engineering, East China University of Science and Technology, Shanghai 200237, P. R. China.}
 \affiliation
{Institute for Theoretical Physics, Utrecht University, Princetonplein 5, 3584 CC Utrecht, The Netherlands.}
\author{Wei Zhong}%
 \email{zhongwei2284@hotmail.com.}
\affiliation{ 
Minjiang Collaborative Center for Theoretical Physics, College of Physics and Electronic Information Engineering, Minjiang University, Fuzhou 350108, P. R. China.}
\affiliation{ 
Department of Information and Computing Sciences, Utrecht University, Princetonplein 5, 3584 CC Utrecht, The Netherlands.}

\date{\today}

\begin{abstract}
Passive particle transport through narrow channels is well studied, while for active particle system, it is not well understood. Here, we demonstrate the active control of the transport through a nanopore via mean-field analysis and molecular dynamics simulations. We prove that the active force enhances the transport efficiency with an effective diffusion coefficient $D_{\text{eff}}=D_t (1+Pe^2/6)$, where $D_t$ is the translational diffusion coefficient, and $Pe$ is the P\'{e}clet number that determines the strength of the active force. For the number of particles inside the channel, it experiences subdiffusion at short times and then turns to normal at longer times. Finally, we extend our research for several sinusoidal shapes of the channel surface. More particles are trapped in the channel if the roughness of the channel surface is increased, resulting in fewer particles are transported from one side of the channel to the other. 
\end{abstract}

\maketitle

\section{Introduction}
The transport of molecules through a micro- or nanopore
 plays a vital role in many biological and chemical systems \cite{mari}. 
 The translocation dynamics of DNA, protein and ions through
 cell membrane are still an ongoing field \cite{jose,amit}, different 
 methods have been used to understand and control those transport processes \cite{ulri,esz,dzu,karo,cheng}. 
 Normally, an external field is required to activate and speed up the
  transport, where the external field can be pressure, temperature or 
  potential differences \cite{mari,man}. However, there is a large group of 
  microorganisms, known as active matter, that makes the external 
  field no longer the only driven source during the transport process. 
 
Active matter, known as the microswimmers or active Brownian 
particles (ABPs), can convert the energy from the environment into 
its own self-propelled velocity \cite{poma1,clem1,srir1}. Recently, active matter attracts 
a lot of attention. After the first well-known model to describe 
the flocking of the birds, i.e., the Vicsek model \cite{vicsek}, 
the field of active matter receives a number of important progresses. 
One of the most impressive progress is the motility-induced phase 
separation (MIPS) \cite{cates1,cates2,pas}, which helps us understand 
 the phase behavior of the ABPs, although the universality 
class at its critical point is still under debate \cite{jona,benj}. 
Besides, many researchers study the dense phase of 
the ABPs, with the glass transition and hexatic 
phase \cite{sid,pas2,ludo1,ludo2,ludo3,nat}. However, in real experiments, 
it is hard to squeeze the active particle into very dense phase, 
the mixture of the active and passive particles brings a possible 
solution \cite{cates3,juli,sho}. 

Most studies about active matter focus on the bulk properties, obtained under periodic 
boundary conditions, few has paid attention to the transports of active matter under confinement. 
The active particles can be trapped or sorted by the solid boundary, the concentration gradient,
or the obstacles \cite{kais,vin,ran,olek}, and the motion of the active particles 
can be rectified by a pattern microchannel \cite{kou}. When active 
particles are moving inside a channel \cite{cos,dhar,ai1,ai2}, it was found that the week external field has strong influence on velocity profiles and particle flow. Even though the 
transport dynamics of the active Brownian particles passing through 
a nanopore is not well understood.

In this paper, we use molecular dynamics (MD) simulations to investigate the 
transport dynamics of the active Brownian particles passing through 
a nanopore. We first demonstrate the mean-field analysis of the 
transport dynamics, revealing that for large self-propelled speed, the active force accelerates 
the transport with an effective diffusion 
coefficient $D_{\text{eff}}= D_t (1+Pe^2/6)$, where $D_t$ is the translational diffusion coefficient, and $Pe$ is 
the P\'{e}clet number that determines the strength of the active force. 
Then we compare our MD results to the mean-field analytical solution, it is found that they are consistent. The mean-square 
displacement (MSD) of the particle number in the channel is also 
obtained, indicating that the diffusion of particle in the channel follows 
anomalous diffusion at short times, then turn to normal diffusion 
at long times. Finally, the effect of the channel surface roughness on the transport is studied. In contrast with the 
straight channel, more particles are willing to stay in the channel 
if the roughness increases, resulting in slower passing through 
dynamics of the active particles. 

\section{Overdamped active Brownian particles}

We consider a two dimensional (2D) system with two rectangle 
chambers $A$ and $C$ (As shown in Fig. 1), whose sizes are $L_x=50 \sigma$ and $L_y=20 \sigma$, 
connected by a narrow channel $B$ of length $l=10\sigma$ and width $d=3 \sigma$, 
where $\sigma$ is the nominal particle diameter. The boundary is built by the active Brownian particles with fixed position.

 The particles in the system interact with each other via the Weeks-Chandler-Anderson (WCA) potential, 
i.e., $U(r)=4\epsilon [(\sigma/r)^{12}-(\sigma/r)^6]+\epsilon$ if $r\leq 2^{1/6}$, 
and zero otherwise. Here, $\epsilon$ represents the interaction strength, and $r$ 
is the center-to-center distance between two particles. We use $\epsilon=1/\beta=k_B T$ 
during our simulation, where $\beta$ is the inverse thermal energy, $k_B$ is 
the Boltzmann constant, and $T$ is the temperature of the system. The model can be described as the overdamped Langevin 
equation \cite{redner,fily,redner2},
\begin{equation}
 \begin{array}{l}
  \dot{\mathbf{r}_i}=D_t\beta [\mathbf{F_i}+F_p \mathbf{\hat{\nu}_i}]+\sqrt{2D_t}\mathbf{\eta}_t \\  
  \dot{\theta}_i=\sqrt{2D_r} \eta_r
\end{array}
\end{equation}
where $\mathbf{r}_i(t)$ and $\theta_i(t)$ are the position and orientation of 
the $i$th particle at time $t$. $\mathbf{F_i}$ is the total excluded-volume 
repulsive force on the $i$th particle, which is given by the WCA potential. $F_p$ 
is a constant self-propulsion force on particle $i$, and the 
direction function $\mathbf{\hat{\nu}_i}=(\cos(\theta_i),\sin(\theta_i))$. 
$D_t$ and $D_r=3D_t/\sigma^2$ represent the translational and rotational 
diffusion coefficient, respectively. The translational and the rotational time scales are 
$\tau_t = \sigma^2 /2D_t$ and $\tau_r = 1/2D_r$. $\mathbf{\eta}_t$ 
and $\eta_r$ are two Gaussian white noise terms with $\langle \eta_t \rangle=0$, 
$\langle \eta_r \rangle=0$ and $\langle \eta_t(t)\eta_t(t')\rangle=2D_t \delta(t-t')$, 
$\langle \eta_r(t)\eta_r(t')\rangle=2D_r \delta(t-t')$. The P\'{e}clet number, defined as the ratio of advective
and diffusive transport rates, is defined as $Pe=v_p \tau_t/\sigma$, where 
$v_p=D_t \beta F_p$ is the self-propelled speed of an individual particle. 

At the beginning of the molecular dynamics (MD) simulations, we place $N$  
particles uniformly distributed in the top chamber (Fig.1), then evolve the system. At each time, we measure the numbers of particles in different regions, i.e., $N_A$, $N_B$, and $N_C$ are the particle numbers in the top chamber, channel, and bottom chamber, respectively. The total number of 
the particles is $N\equiv N_A+N_B+N_C$.

 The molecular dynamics (MD) simulations used here are coded in {\it c programming} and the time unit used here refer to one move per particle within a short time $dt=0.00001$. Running about 25 independent samples provide us with fairly accurate values of the particle numbers in different position and time.

\begin{center}
\begin{figure}[htbp]
\includegraphics[width=0.8 \linewidth]{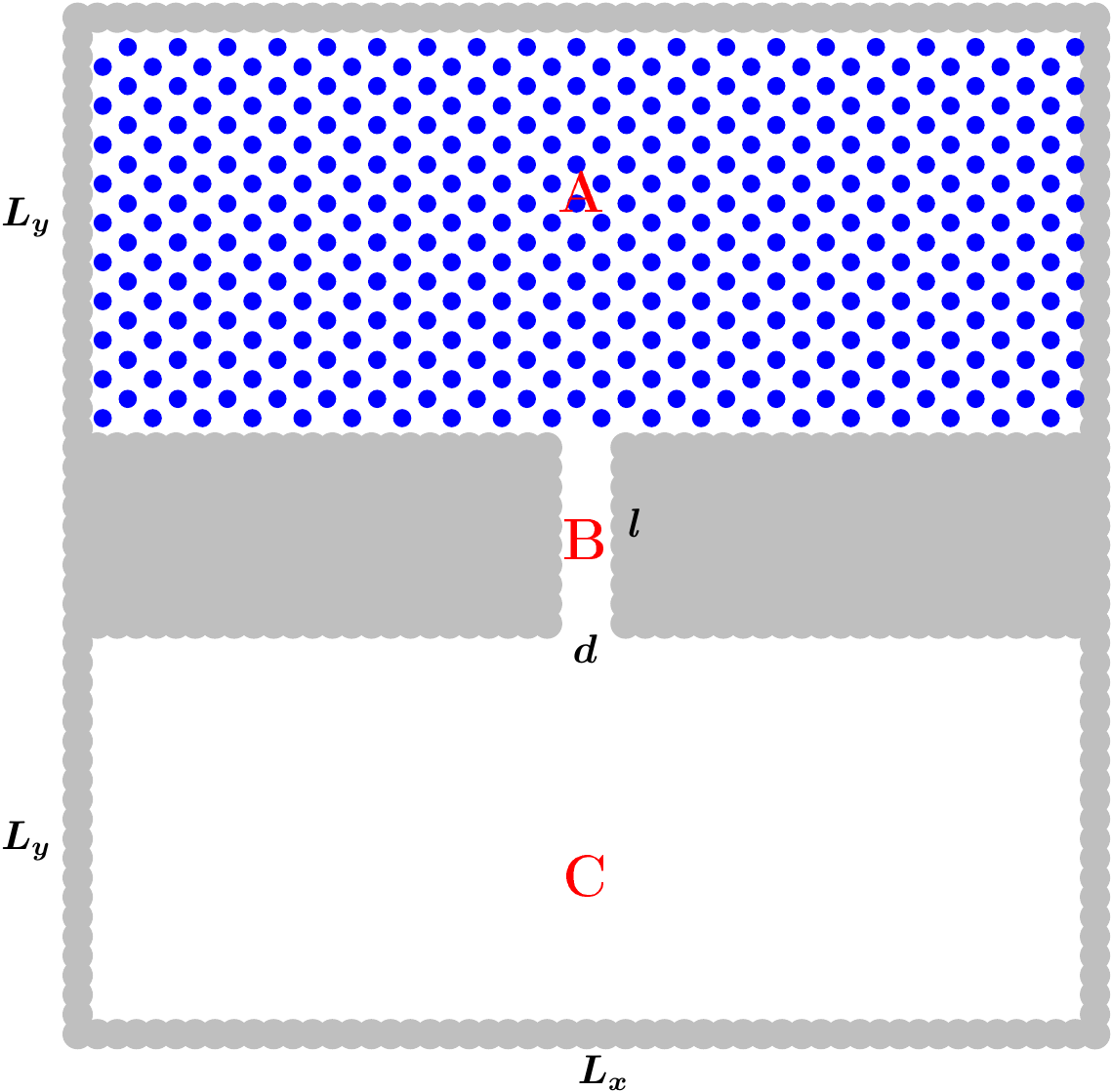}
  \caption{\footnotesize The schematic diagram of the $2D$ system. Two rectangle 
  chambers, with their sizes $L_x=50 \sigma$ and $L_y=20\sigma$, 
  are connected by a narrow channel of length $l=10\sigma$ and width 
  $d=3 \sigma$. All active particles are placed uniformly in the upper chamber at $t=0$.}
  \label{system} 
\end{figure}
\end{center}

\section{Nano-scale transport of ABPs}
\subsection{Mean-field theory understanding}

Supposing we set a uniform distribution of the initial
angles $\theta(0)=0$, the mean-square displacement from the initial position 
$r(0)$ evolves as \cite{howse,locate}

\begin{equation}
\begin{aligned}
 \langle [r(t)-r(0)]^2 \rangle&=4D_t t+\frac{v_p^2 \tau_r^2}{2}[\frac{2 t}{\tau_r} +e^{-2t/\tau_r}-1] \\
 &\approx 4 D_{\text{eff}} t.
\end{aligned}
\end{equation}

When $t\gg \tau_r$, the effective diffusion coefficient is expressed as
\begin{equation}
\begin{aligned}
D_{\text{eff}}&= D_t+\frac{v_p^2 \tau_r}{4} 
\\&= D_t(1+Pe^2 \frac{D_t}{2 \sigma^2 D_r})\\
&= D_t(1+\frac{Pe^2}{6}).
\end{aligned}
\end{equation}

The diffusion equation for the density of active Brownian particles $c(y,t)$ 
can be written as 
\begin{equation}
\partial_t c(y,t) = \partial_y D_{\text{eff}} \partial_y c(y, t),
\label{partial}
\end{equation}
 
Base on the analytic solution of Eq. \ref{partial} from refs. \cite{kond1,kond2} and the similar behavior found in the ions transport in supercapacitors\cite{kon,tang,bi},
one finds that the particle number $N-N_A$ increases linearly in time at early 
times:
\begin{equation}
N-N_A\sim t
\label{pe0}
\end{equation}
 Then it experiences a square-root behaviour with a short time range, i.e.,
\begin{equation}
N-N_A\sim \sqrt{D_{\text{eff}}}\, t.
\label{pe}
\end{equation}

Finally, $N-N_A$ increases exponentially in time as
\begin{equation}
N-N_A\sim 1-\exp(-\sqrt{D_{\text{eff}}}\,t/\tau).
\label{pe2}
\end{equation}
where $\tau$ is a constant.

 The mean-field analysis indicates that the active force accelerates the 
 transport with a prefactor linear in $\sqrt{D_{\text{eff}}}$. 

\subsection{Dynamics from MD simulations}

As discussed through the mean-field analysis, we expect that the active force will accelerate the transport dynamics. In order to determine the acceleration of the dynamics by the active force numerically, two different sets of simulations have been performed using the MD algorithm. In the first set, the P\'{e}clet number is fixed at $Pe=200$. Since we only focus on dilute 
situation, the initial number of particles on the top chamber is 
set to be $N=400$, $225,$ and $100$. In the other set, under the constrain that the number of particles is $N=400$, three different P\'{e}clet 
numbers $Pe=50$, $100$ and $200$ are employed during the simulations.

\begin{figure*}[t!]
\centering
\includegraphics[width=0.85 \linewidth]{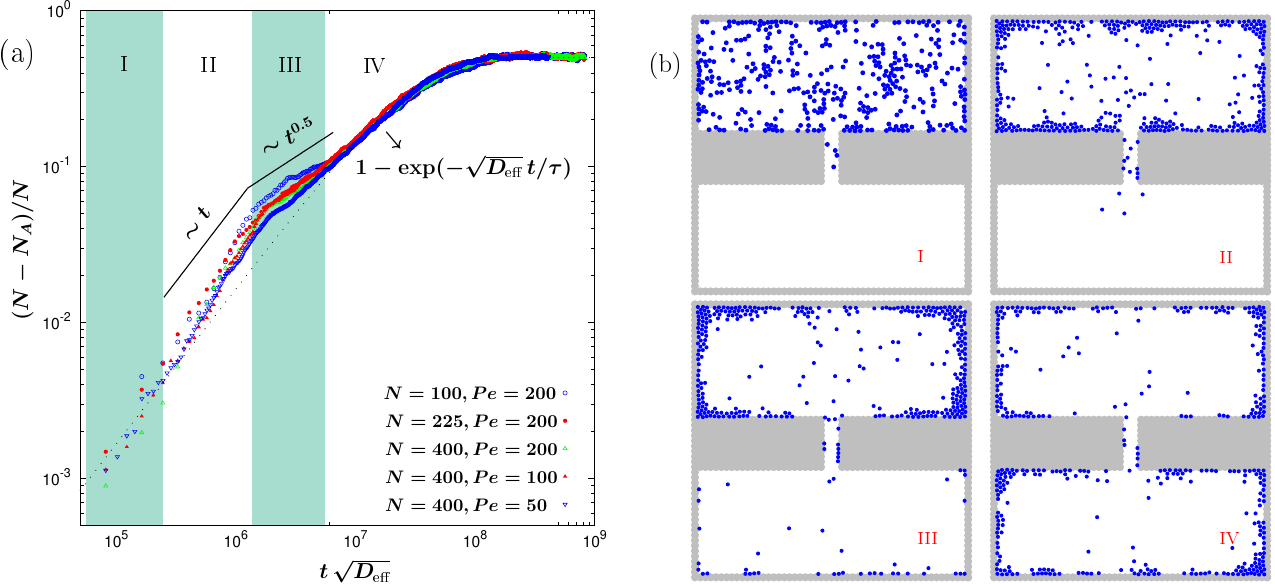}
  \caption{\footnotesize  (a) Dynamics of the particle number $N-N_A$ for different $N$ and $Pe$. At very early times, particles are still exploring the area $A$ 
  and $B$. Then some of the particles begin to escape from the channel and move to $C$. However, they do not feel the space confinement of $C$, 
  indicates that $N-N_A$ is linear in time. 
  At the intermediate times, the space confinement in $C$ slows down the speed for a particle passing from $A$ to $C$. 
  At this stage, $N-N_A\sim t^{0.5}$. Finally, particle number in $A$ and $C$ are near the steady-state, 
  their behavior becomes exponential. (b) Typical snapshots for different time periods.}
  \label{dyna} 
\end{figure*}

The resuts shown in Fig. \ref{dyna}(a) suggest that indeed the active force accelerates the 
dynamics with a prefactor linear in $\sqrt{D_{\text{eff}}}$, i.e., $(N-N_A)/N\sim f(t \, \sqrt{D_{\text{eff}}})$. After a noisy initial part, $N-N_A$ increases linear in time. Then it turn to a square-root growth for a short time range. For longer times, $N-N_A \sim 1-\exp(-\sqrt{D_{\text{eff}}}\,t/\tau)$. These results are in agreement with the mean-field analysis (Eq. (\ref{pe0}) - (\ref{pe2})). 

The snapshots for different stages give more details about the transport process:

{\textbf{Very early stage I}}: As shown in fig.\ref{dyna} (b) I, at the beginning of the simulation, the particles are 
exploring the top chamber and the channel, none of them moved to the bottom 
chamber yet. 

{\textbf{Linear regime II}}: After stage I, some of the particles begin 
to reach the bottom chamber. However, they still do not feel the spatial 
confinement of the bottom chamber, therefore particles are continuously entering into the bottom chamber, i.e., $N-N_A\sim t$. 

{\textbf{Square-root diffusive regime III}}: As more particles enter into the bottom chamber, the spatial confinement becomes important, which slows down the particles moving from top chamber to bottom chamber, 
resulting in $N-N_A\sim \sqrt{t}$. 

{\textbf{Exponential regime IV}}: Finally, the competition of the numbers of particles 
in both chambers indicates $N-N_A$ increases exponentially.

\subsection{Dynamcis of the in-pore particles}

To understand the dynamics of the particles in the channel, we define the 
mean-square deviation (MSD) of the in-pore particle number $N_B$ as 
\begin{equation}
\langle \Delta N_B(t)^2\rangle=\langle [N_B(t)-N_B(0)]^2\rangle
\end{equation}

In our MD simulations, when the system reaches its steady state, we generate 
a long enough time series of $N_B(t)$, from which we calculate the MSD of $N_B$. 

\begin{figure}[htp]
\centering
\includegraphics[width=0.8 \linewidth]{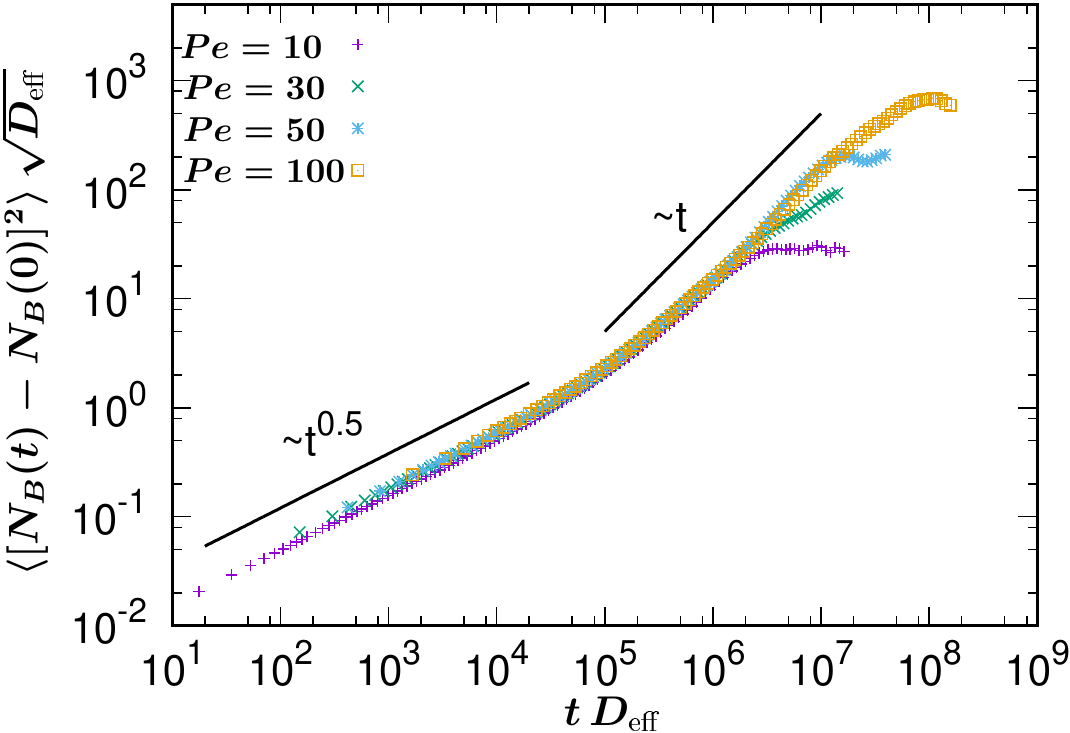}
  \caption{\footnotesize The mean-square deviation of the particle number in the channel 
  $\langle \Delta N_B(t)^2\rangle$ for different $Pe$. For $t\, D_{\text{eff}} < 10^5$, 
  it goes as $\langle \Delta N_B(t)^2\rangle \sim t^{0.5}$. After that, $\langle \Delta N_B(t)^2\rangle \sim \sqrt{D_{\text{eff}}}\,t$.}
  \label{fig_msd} 
\end{figure}

As shown in Fig. \ref{fig_msd} that the MSD of $N_B$ behaviors as 
$\langle \Delta N_B(t)^2\rangle\, \sqrt{D_{\text{eff}}} \sim f(t D_{\text{eff}})$. At short 
times ($t\lesssim 10^5/D_{\text{eff}}$), $\langle \Delta N_B(t)^2\rangle \sim t^{0.5}$, 
which is a regime that the behavior of $\langle \Delta N_B(t)^2\rangle$ is 
independent of the active force and it is identical to the diffusion of particles 
in the single-file \cite{hahn,lutz}. After that the number of particles in the 
channel diffuse normally.  

In summary, for most existed systems, an external field is necessary to drive the particles to pass 
through the narrow channel. However, in nature, especially in many biological 
systems, some of the particles or microorganisms have a self-propelled force, and 
for those kind of systems, the active force itself dramatically improves the 
efficiency for particles to pass through a narrow channel. As we can see from both the mean-field and simulation results that the active force enhances the transport with an effective diffusion coefficient $D_{\text{eff}}$. Besides, the number of particles inside the channel experiences anomalous diffusion at short times. 

\begin{figure*} [htp]
\centering
\includegraphics[width=0.85 \linewidth]{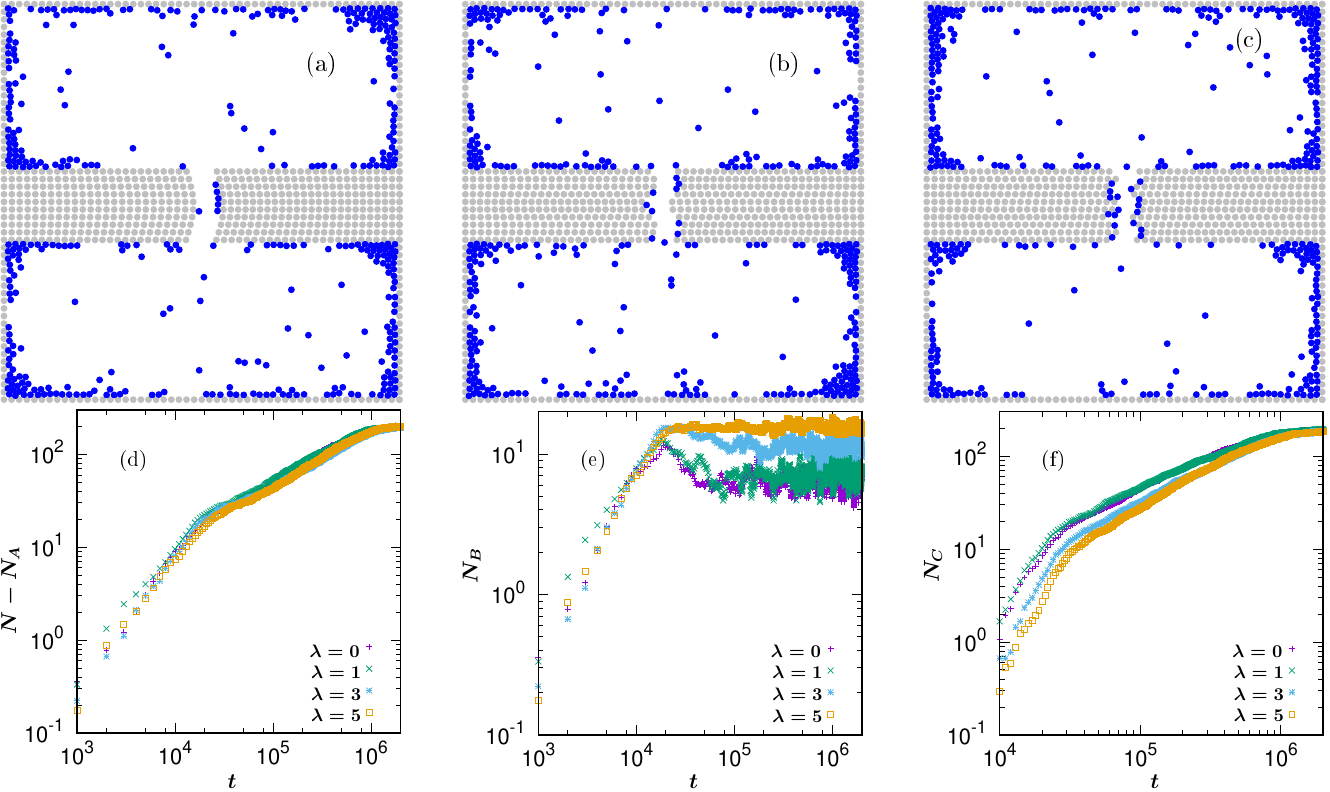}
  \caption{\footnotesize (a)-(c) Snapshots when systems reach the steady-state 
  for different roughness. Here $Pe=200$, and the shape of the channel is 
  expressed as $\alpha \sin(\lambda (y-L_y)/\pi)$, with $\alpha=1$ and 
  $\lambda=1,3,5$ for (a)-(c), and when $\lambda=0$, the channel is a straight 
  channel as discussed in the previous section. (d)-(f) The effect of the roughness on the transport dynamics of the active Brownian particles passing through the narrow channel. The influence of the roughness to the number of particles left the top chamber is tiny, however, 
  more particles would like to stay in the channel because of the roughness, so that the speed for particles go to the bottom chamber is decreasing.}
  \label{rough} 
\end{figure*}

\section{Affecting transport by channel roughness}

In the previous section, we only focus on the straight channel. However, in 
most situations, the shape of the channel is not necessary to be straight. Here, 
the shape of the channel is modified as
\begin{equation}
W(y)=\alpha \sin \left[\lambda \frac{y-(L_y+1)\sigma}{\pi}\right] \quad \text{for} \quad L_y\leq y<L_y+l,
\end{equation} 
where $\alpha$ is set to be unity, and $\lambda$ is a roughness parameter. 
When $\lambda$ is not equal to zero, it changes the roughness of the channel surface. 

Three different $\lambda$ (with $\lambda=1,3$ and $5$) are considered. The particles 
are more willing to stay in the corner, therefore if we increase the roughness 
of the channel, more particles will be trapped in the channel (See TABLE I and Fig.4 (e) for more details.) and the speed for 
particles to move to the bottom chamber decreases.

\begin{table*} 
\centering
\begin{tabular}{|m{1.5cm}|m{1.5cm}|m{1.5cm}|m{1.5cm}|m{1.5cm}|m{1.5cm}|m{1.5cm}|}
\hline
$t$ & $5000$ & $10000$ & $50000$ & $100000$ & $500000$ & $1000000$\\
\hline
$N_B(\lambda=0)$ & 0.92(2) & 1.92(1) & 1.96(2) & 1.52(2) & 1.80(2) & 1.38(3)\\
\hline
$N_B(\lambda=1)$ & 2.12(6) & 7.88(6) & 8.44(10) & 7.22(12) & 6.00(12) & 8.00(10)\\
\hline
$N_B(\lambda=3)$ & 3.00(10) & 7.44(12) & 13.22(8) & 11.00(8) & 11.22(6) & 10.44(8) \\
\hline
$N_B(\lambda=5)$ & 2.82(5) & 7.05(6) & 15.23(4) & 15.52(4) &  14.70(6) & 15.17(3) \\
\hline
\end{tabular}
\caption{The average values of the number of
active particles moving within the channel, i.e., $N_B$, at some particular time for different roughness parameters $\lambda=0, 1, 3, 5$. Here, the P\'{e}clect number is set to be $Pe=200$, the pore length and size are $l=10\sigma$ and $d=4\sigma$. At the initial time, $400$ active particles are uniformly distributed on the top chamber.}
\label{tab11}
\end{table*}

Fig. \ref{rough}(d) shows that for different roughness, the dynamics of $N-N_A$ is 
not affected dramatically. However, Fig. \ref{rough}(e) confirms that more particles stay in the channel by increasing the roughness, leading to  less number of
particles move to the bottom chamber than the straight channel situation (As shwon in Fig. \ref{rough}(f)).
  
\section{Conclusion}
In this paper, we consider a two dimensional (2D) system with two rectangle chambers 
connected with a narrow channel. We distribute a few hundred of overdamped active 
Brownian particles in the top chamber, then evolve the system. 

By using the mean-field analysis and MD simulations, we find that the number of particles $N-N_A$ goes as $(N-N_A)/N\sim f(t \, \sqrt{D_{\text{eff}}})$. At short times, $N-N_A$ increases linear in time. Then after a slower dynamics regime, 
i.e., $N-N_A\sim \sqrt{t}$, the dynamics become exponential. In the meanwhile, at short 
times, the number of the particles in the channel experience anomalous diffusion 
$\langle \Delta N_B(t)^2\rangle\sim \sqrt{t}$, then turn to normal diffusion at $t\gtrsim 10^6/D_{\text{eff}}$.  

Finally, we change the roughness of the channel to be a {\it sinusoidal} shape. It is found 
that the straight channel is the most effective shape for active particles to 
pass through. Since active force is ubiquitous in many biological systems, when 
we try to understand the mechanism for microorganisms passing through a nanopore, 
the effect from the active force should be considered. 


\section*{Acknowlegement \label{sec5}}
This research was sponsored by the EU-FET project NANOPHLOW (REP-766972-1), the National Natural Science Foundation of China (Nos. 91834301 and 22078088), and Shanghai Rising-Star Program (21QA1401900). The authers would like to thank Y. K. Zhu for his help in some simulations. We kindly thank J. de Graaf, R. van Roij, G. T. Barkema, and P. Huang for helpful discussions.

\section*{Data Availablity}
The data that support the findings of this study are available from the corresponding author upon reasonable request.







\end{document}